\documentstyle[aps,multicol,epsfig,amsmath,amssymb]{revtex}

\begin{document}
\title{A source of cold atoms for a continuously loaded magnetic guide}

\author{C.F. Roos, P. Cren, J. Dalibard and D. Gu\'ery-Odelin}

\address{Laboratoire Kastler Brossel\cite{LKB}, Ecole Normale Sup\'erieure}
\address{24, Rue Lhomond, F-75231 Paris Cedex 05, France}

\date{\today}

\maketitle

\begin{abstract}
We present an intense source of $^{87}$Rb atoms that has been set up to produce a continuous, slow
and cold beam in a magnetic guide. It consists of a two-dimensional magneto-optical trap whose
cooling laser power is provided by a master-oscillator tapered-amplifier system. This trap produces
an atomic beam with a flux of over $10^{10}$ atoms/s and a mean velocity of 40 m/s. The beam is
recaptured by a second trap whose purpose consists in reducing the beam's velocity, in further
cooling the atoms and injecting them into the magnetic guide. The article focusses on the first stage
of the process described above.
\end{abstract}


\begin{multicols}{2}

A spectacular challenge in the  field of Bose-Einstein condensation consists in the achievement of a
continuous beam operating in the quantum degenerate regime. This would be the matter wave equivalent
of a cw monochromatic laser and it would allow for unprecedented performances in terms of
focalization or collimation. In \cite{ScienceK02}, a continuous source of  Bose-Einstein condensed
atoms was obtained by periodically replenishing a condensate held in an optical dipole trap with new
condensates. This kind of technique raises the possibility of realizing a continuous atom laser.  An
alternative way to achieve this goal has been studied theoretically in \cite{Mandonnet00}.
 A non-degenerate, but already slow and cold beam of particles, is injected into a magnetic guide
\cite{Schmiedmayer95,Denschlag99,Goepfert99,Key00,Dekker00,Teo01,Sauer01,Hinds99} where transverse
evaporation takes place. If the elastic collision rate is large enough, efficient evaporative cooling
 can lead to quantum degeneracy  at the exit of the guide.  The condition for reaching
degeneracy with this scheme can be formulated by means of three parameters: the length $\ell$ of the
magnetic guide on which evaporative cooling is performed, the collision rate $\gamma$ at the
beginning of the evaporation stage, and the mean velocity $v_{\rm b}$ of the beam of atoms. Following
the analysis given in \cite{Mandonnet00}, one obtains
\begin{equation}
\frac{\gamma\ell}{v_{\rm b}}\gtrsim 500 \; . \label{cond1}
\end{equation}
 Physically, (\ref{cond1}) means that each atom has to undergo
at least 500 elastic collisions during its propagation through the magnetic guide.

Our experiment aims at implementing this scheme for a beam of $^{87}$Rb atoms. Its success relies
therefore upon two preliminary and separate accomplishments. First, one has to build an intense
source of cold atoms, with the lowest possible mean velocity. Second, one has to inject the atomic
beam produced by this source into a long magnetic guide with minimal transverse and longitudinal
heating. In our experiment, we subdivide the first task into the production of an intense atomic beam
which is only subsequently further slowed down and cooled.

We generate a high flux of atoms by means of a two-dimensional magneto-optical trap (2D-MOT). Atoms
of the beam produced by this source are recaptured by a second atom trap whose purpose consists in
injecting the atoms into the magnetic guide with a very low velocity ($v_{\rm b}\simeq\,1$m/s) that
can be chosen at will. The present paper characterizes our 2D-MOT and is organized as follows: the
principle of operation of the experiment is presented in section one. The second and third part of
the paper respectively describe the experimental setup and the results that we have obtained so far.

\section{Principle of operation}
In order to inject atoms into the magnetic guide, we use a magneto-optical trap (MOT) based upon four
laser beams in a tetrahedral configuration superimposed with a magnetic two-dimensional quadrupole
field. The setup consists of a moving molasses \cite{MM} in the longitudinal direction combined with
cooling and confining forces in the transverse directions and will be called {\it injecting MOT} in
the following. A detailed description can be found in \cite{Cren}. The injecting MOT allows to
capture atoms entering the trapping volume with a velocity $v<v^\star$, where $v^\star$ is a function
of the available laser power, and to subsequently slow the atoms down to the final velocity $v_{\rm
b}$ which only depends on the relative frequencies of the laser beams. In this way we produce an
atomic beam with a temperature on the order of 50 $\mu$K and a velocity in the range of 30 cm/s to 3
m/s.

If the injecting MOT is loaded from a rubidium background gas with vapor pressure $p$, it captures
atoms at a rate proportional to $p$. However, collisions between captured and thermal background
atoms will limit the flux $\Phi$ of outgoing atoms to
\begin{equation}
\Phi \propto p\;\exp(-\frac{\alpha p}{v_{\rm b}})\;, \label{flux}
\end{equation}
where $\alpha$ depends on the characteristic length of the trap and the collision cross section. This
process sets an upper limit $\Phi_{{\rm max}}=\max_{\{p\}}(\Phi)$ to the flux that can be achieved.
It decreases as $\Phi_{{\rm max}}\propto v_{\rm b}$ with decreasing velocity $v_{\rm b}$. Since
$\gamma\propto\Phi/v_{\rm b}$, it also limits the collision rate in the magnetic guide provided that
the temperature and the confining force of the guide are unchanged. In our previous setup
\cite{Cren}, we were able to produce a flux of $10^9$ atoms/s at $v_{\rm b}=2.5$ m/s and a pressure
$p=4\times 10^{-8}$ mbar. In order to overcome this limitation, we have to operate the injecting MOT
in an ultra-high vacuum chamber.

For this purpose, many techniques have been developed. Thermal beams can be slowed down by means of a
Zeeman slower \cite{Zeeman}. Alternatively, a collimated beam of atoms can be obtained by a
vapor-loaded MOT with a leak at its center. For instance, one can drill a hole in one of the mirrors
of a MOT \cite{Lu96,Walraven98}), or use a pyramidal mirror structure with a hole at its vertex
\cite{Lee96,Williamson98,Arlt98}, or use an extra pushing beam, which destabilizes the MOT at its
center \cite{Wohlleben01,Cacciapuoti01}. A third method for producing a cold beam of atoms relies on
a 2D-MOT \cite{Walraven98,Pfau02}. All these sources produce a relatively intense beam (from 10$^6$
to several $10^{10}$ atoms per second), with an average velocity $\bar v$ between 10 and 50~m/s. In
our setup, we have implemented the 2D-MOT scheme. We note that the relation (\ref{flux}) also applies
to the source producing the atomic beam. However, it is much less restrictive since the atomic
velocities at its exit are orders of magnitude higher than in the injecting MOT.

\section{Experimental setup}
\begin{figure}[hbt]
\begin{center}
  \epsfig{file=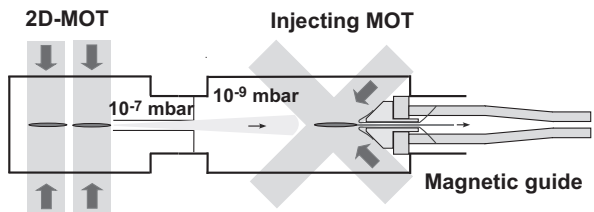,width=.9\linewidth}
  \caption{\label{setup}
  Schematic drawing of the experimental setup. Atoms are captured from the background gas by the
2D-MOT which creates an atomic beam that is (partly) recaptured by the injecting MOT. The latter
injects atoms into the magnetic guide with well-controlled velocity.}
\end{center}
\end{figure}

A schematic drawing of the experimental setup is shown in fig.~\ref{setup}. The first MOT is located
in a rectangular quartz cell (80 mm $\times$ 45 mm $\times$ 45 mm) connected to a Rubidium reservoir
by a valve (not shown in fig.~\ref{setup}) and joined to the main part of the vacuum system by a
differential pumping tube. It is a vapor-loaded two-dimensional MOT based on a design that has been
characterized in detail in \cite{Walraven98,Pfau02}. For a thermal atom to be captured by the MOT,
firstly it needs to have a transverse velocity smaller than the MOT's capture velocity. Secondly, the
atom's longitudinal velocity $v_{\parallel}$ has to be low enough so that the transit time
$\tau=l/v_\parallel$ ($l$ being the length of the MOT) is sufficiently long for the atom to be
transversally trapped and cooled. In this way the 2D-MOT produces two beams of atoms which leave the
trap on axis in opposite directions.

Two pairs of elongated coils in anti-Helmholtz configuration produce a two-dimensional quadrupole
magnetic field whose zero-field line is parallel to the optical table and centered with respect to
the glass cell. Typical field gradients are on the order of 12 G/cm. The cooling laser light for the
MOT is produced by a tapered amplifier providing up to 450 mW of light. The amplifier is injected by
a grating-stabilized diode laser which is red-detuned by $\Delta\simeq -2\Gamma$ from the $5S_{1/2},
F=2 \rightarrow 5P_{3/2}, F=3$ transition. A second grating-stabilized diode laser tuned to the
$5S_{1/2}, F=1 \rightarrow 5P_{3/2}, F=2$ transition is used to injection-lock another laser diode
which provides up to 30 mW of repumping light. The cooling and repumping laser beams are superposed
and split into two beams which are subsequently circularly polarized. After widening the beams by
telescopes to a beam waist of 13 mm, the beams are steered to the quartz cell which they pass
perpendicular to each other and to the zero-field line of the magnetic field. They are then
retro-reflected after having passed another quarter-wave plate, thus providing the third and forth
beam of the MOT. We find that the resulting intensity imbalance between the ingoing and the
retro-reflected beams does not significantly perturb the functioning of the trap. Typically, the trap
is operated at a total rubidium background pressure of $3\times 10^{-7}$ mbar. It is located close to
the entrance of a differential pumping tube, with a diameter $d=7$ mm and a length $l=450$ mm,
capable of maintaining a pressure ratio of more than two orders of magnitude between the first quartz
cell and the cell of the injecting MOT. The total distance between the first and the second MOT is
$L\simeq 800$ mm.

We use two methods to study the properties of the 2D-MOT. By measuring the fluorescence of the
injecting MOT, we obtain a signal which is proportional to the flux of atoms that is produced by the
2D-MOT and recaptured by the injecting MOT. Even though this quantity is the most important parameter
for the next stage of the experiment, it does not allow to characterize the performance of the 2D-MOT
in a quantitative and independent way. Therefore, to gather more information we perform absorption
measurements on the beam of atoms that arrives in the glass cell of the recapturing MOT. For this
purpose, we shine a probe laser beam perpendicular to the atomic beam into the glass cell of the
recapturing MOT. The beam is resonant with the $5S_{1/2}, F=2 \rightarrow 5P_{3/2}, F=3$ transition,
has a diameter of 1 mm and is circularly polarized. A weak magnetic field in the direction of the
probe beam ensures that the atoms crossing the probe beam are optically pumped to the outermost
Zeeman state. By measuring the absorption of the probe beam, we obtain the column density of the
atomic beam as a function of the probe beam position (in the direction perpendicular to the atomic
beam). Information about the velocity distribution of the atoms is obtained by performing
time-of-flight measurements. After suddenly turning off the repumping laser of the 2D-MOT, we measure
the atomic velocity dispersion by monitoring the absorption of the probe beam as a function of time.
Note that this method works since the velocity distribution $\Delta v$ is on the order the mean
velocity $\bar{v}$ and since the distance between the probe beam and the 2D-MOT is large compared to
the size of the MOT. In order to increase the signal-to-noise ratio of the measurement, we dither the
frequency of the probe laser at 100 kHz and use a lock-in amplifier to measure the absorption.

\section{Experimental results}

For the 2D-MOT with circular beam profiles described in the previous section, we find that the number
$N$ of atoms recaptured by the injecting MOT increases approximately linearly as a function of the
cooling laser power $P$  when $P$ was varied between 16 mW and 160 mW per beam. When we reduce the
cooling volume of the trap by introducing knife edges which limit the beam size, $N$ decreases
dramatically in the case where the length of the trap is reduced while it is much less affected if
the transverse beam size is reduced. It has already been observed and explained in \cite{Pfau02} that
elliptical beam profiles are advantageous for maximizing the atomic flux. Here, instead of using
cylindrical lenses to create elliptical beams, we split each of the MOT beams into two beams before
passing them through the telescopes. Each pair of beams is now passed through the glass cell side to
side to each other so as to produce a 2D-MOT with elongated beam profiles (see fig.\ref{setup}). The
laser power of the tapered amplifier can be arbitrarily repartitioned among the four beams, and we
find that a maximum of recaptured atoms is observed for a configuration where the intensity of the
two beams that are closer to the trap exit is lower than the intensity of the other two beams. The
measurements presented in the following have been performed with this optimized configuration.

\begin{figure}[hbt]
\begin{center}
\epsfig{file=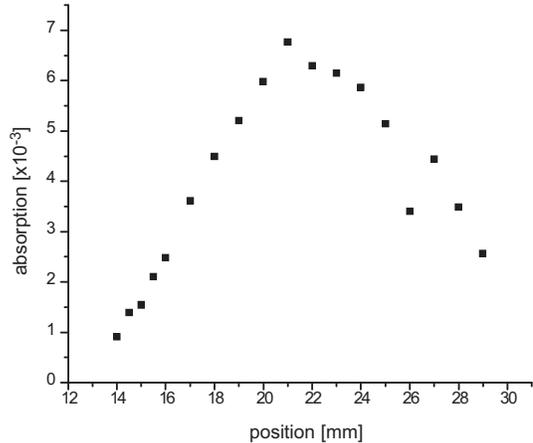, width=7cm,angle=0 } \caption{\label{profile}
 Density profile of the atomic beam at a distance of 80 cm from the source.}
\end{center}
\end{figure}

\begin{figure}
\begin{center}
\epsfig{file=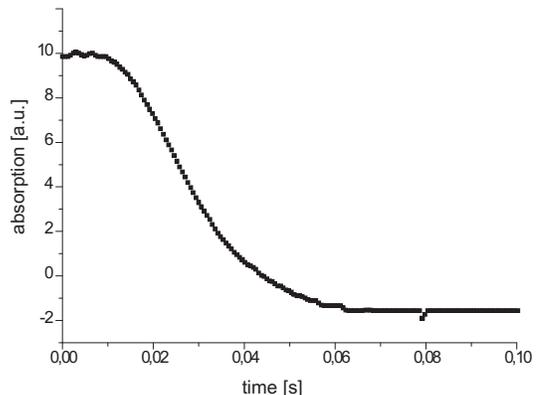, width=7cm,angle=0}
 \caption{\label{signal}
 Absorption signal obtained after switching off the 2D-MOT at t=0.}
\end{center}
\end{figure}

\begin{figure}
\begin{center}
\epsfig{file=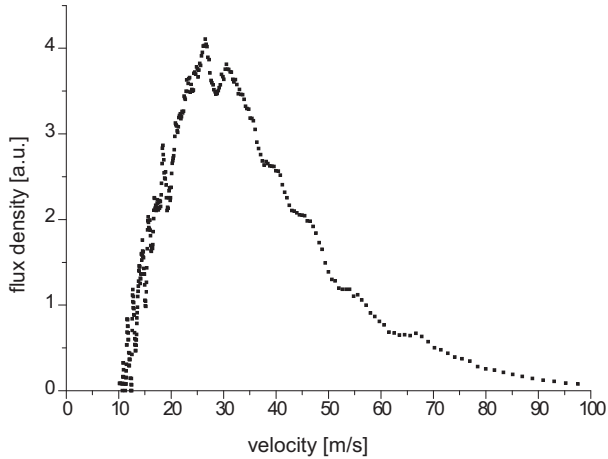, width=8cm,angle=0 } \caption{\label{fluxdensity}
 Atomic flux density $\rho_\Phi(v)$.}
\end{center}
\end{figure}

Absorption measurements of the atomic beam are done at a distance $L\simeq 800$ mm from the 2D-MOT.
The maximum absorption is on the order of $0.7\%$. Even though the absorption signal is small, it
allows to calibrate those measurements that take advantage of lock-in detection. The column density
profile of the atom beam in fig.~\ref{profile} is recorded by scanning the probe beam vertically
(i.~e., in a direction perpendicular to the atomic beam). The full width at half maximum (FWHM) of
about 10 mm is superior to the diameter $d$ of the differential pumping tube so that one might wonder
whether the transverse beam size could be limited by the tube. If the transverse size of the 2D-MOT
was negligible and the spatial density of the atomic flux was homogeneous at the tube exit, the
linear density profile would have the shape of a half circle with diameter $d_c=d\,L/l\approx 12.4$
mm, giving a FWHM of 10.8 mm. A finite transverse trap size and the velocity-dependence of the beam
profile due to gravity contribute to enlarge the beam size.
The velocity distribution of the atomic beam is obtained by rapidly switching off the repumping beam
of the 2D-MOT and measuring the time-dependent absorption signal $s(t)$ as shown in
fig.~\ref{signal}.
The atomic flux density per velocity class $\rho_\Phi(v)$ is related to $s(t)$ by
\begin{equation}
\rho_\Phi(v)\propto t \frac{{\mathrm d} s}{{\mathrm d} t} \;\;\mbox{with}\; t(v)=L/v\; .
\end{equation}
After calibrating the signal and integrating over the linear density profile, we obtain the atomic
flux density which is shown in fig.~\ref{fluxdensity}. The beam is found to have a mean velocity
$\bar{v}=\int {\mathrm d} v\,v\rho_\Phi /\int {\mathrm d} v\,\rho_\Phi = 38$ m/s and a rather large
velocity spread $\Delta v=\langle (v-\bar{v})^2\rangle^{1/2}=17$ m/s. The highest flux that we have
measured was $\Phi\gtrsim 10^{10}$ atoms/s. This number includes atoms with velocities up to 100 m/s.
The experimentally most relevant quantity is the atomic flux that the injecting MOT is able to
recapture. The capture velocity that can be achieved strongly depends on the trap geometry as well as
the laser power available for the injecting MOT and will typically lie in a range between 25 m/s and
40 m/s. The fraction of atoms with a velocity below 35 m/s is 50\%, which corresponds to a useful
flux of $5\times 10^9$ atoms/s.

In this article, we have reported on the performance of a 2D-MOT similar to the one described in
\cite{Walraven98,Pfau02}. In our setup, we use two adjacent 2D-MOT which allows to maximize the
atomic flux by redistributing the available laser power among the two traps. This source has been
developed for loading a moving molasses MOT which serves to inject a cold and slow atomic beam into a
magnetic guide.

\section*{acknowledgements}
This work was supported by the Bureau National de la M\'etrologie, the D\'el\'egation G\'en\'erale de
l'Armement, the Centre National de la Recherche Scientifique and the R\'egion Ile de France. C. F.
Roos acknowledges support from the European Union (contract HPMFCT-2000-00478).

\end{multicols}
\end{document}